\documentstyle[aps,psfig]{revtex}
\newcommand{\s}{\mathrm}
\newcommand{\bd}{\mathbf}
\newcommand{\ra}{\rightarrow}
\newcommand{\lr}{\leftrightarrow}
\newcommand{\mn}{\mu \nu}
\newcommand{\be}{\begin{equation}}
\newcommand{\ee}{\end{equation}}

\newcommand{\ba}{\begin{eqnarray}}
\newcommand{\ea}{\end{eqnarray}}
\newcommand{\bef}{\begin{figure}}
\newcommand{\eef}{\end{figure}}

\newcommand{\ep}{\epsilon}
\newcommand{\lda}{\lambda}
\newcommand{\rpp}{\rho \pi \pi}
\newcommand{\opr}{\omega \pi \rho}

\begin{document}

\begin{center}
{\Large{Unstable particles in matter at a finite temperature:
the rho and omega mesons}\\
}
\vskip .2in
Jan-e Alam$^1$, Sourav Sarkar$^1$, Pradip Roy$^1$,
Binayak Dutta-Roy$^2$,\\
and\\
Bikash Sinha$^{1,2}$\\
\it{$^1$Variable Energy Cyclotron Centre,
     1/AF Bidhan Nagar, Calcutta 700 064
     India}\\
\it{$^2$Saha Institute of Nuclear Physics,
           1/AF Bidhan Nagar, Calcutta 700 064
           India}

\end{center}
\vskip .2in
\parindent=20pt
\vskip 0.1in
\begin{abstract}
Unstable particles (such as the vector mesons) have an important 
role to play in low mass dilepton production resulting 
from heavy ion collisions and this has been a subject of several
investigations. Yet subtleties, such as the implications
of the generalization of the Breit-Wigner formula for nonzero
temperature and density, e.g. the question of collisional broadening, 
the role of Bose enhancement, etc., the possibility of the
kinematic opening (or closing) of decay channels due
to environmental effects, the problem of double counting 
through resonant and direct contributions, are often given 
insufficient emphasis. The present study attempts to point out
these features using the rho and omega mesons as illustrative 
examples. The difference between the two versions of the
Vector Meson Dominance Model in the present context is also
presented. Effects of non-zero temperature and density,
through vector meson masses and decay widths, on dilepton spectra 
are studied, for concreteness  within the framework of a Walecka-type model, 
though most of the basic issues highlighted apply to other scenarios as well.
\end{abstract}



\section*{I. Introduction}
Heavy ion collisions at high energies produce matter far above 
the ground state providing thereby a rich arena for the study of hot
hadronic matter, possibilities of chiral symmetry restoration,
transition to a quark gluon plasma etc~\cite{Hwa}. 
However, hadronic signals are generally 
unsuitable for the task of uncovering the information 
on the underlying occurrences, since the history of 
strongly interacting particles entail layers of
complicated dynamics which mask the basic issues.
As such, electromagnetic signals,
as manifested through emitted photon and dilepton spectra, are 
relatively cleaner, since electromagnetic quanta couple but weakly to
hadronic matter. Final spectra exhibit resonance structures, which,
in the low mass region, include the rho and omega mesons. Consequently,
details of their creation, propagation and decay 
in the medium (or outside) are of paramount
importance for the analysis of the resultant spectra observed.
The general framework for such an investigation  has been provided by 
Weldon~\cite{Weldon} through a beautiful and lucid exposition on the 
Breit-Wigner (BW) formula at non-zero temperature and density.

Accordingly section II, devoted to the underlying principles,
begins with a quick review of the main results of Weldon's paper,
followed by a  brief discussion of the basic ideas of vector meson dominance
(VMD)~\cite{Sakurai} which enables a phenomenological introduction of the
coupling of photons to hadrons and hence to lepton pairs.
We close our presentation of basics 
through an outline of a Walecka-type model~\cite{vol16} 
which is merely used, in the present context, to provide a
setting wherein we can formulate the hadronic scenario allowing us 
to estimate the differences that can accrue if insufficient emphasis
is placed on the subtleties. We go on in section III to present 
the necessary ingredients to evaluate the dilepton emission 
rate from vector meson decays and the pion annihilation process
in hot and dense hadronic surroundings. The last 
section is devoted to a discussion of results and
conclusions.

\section*{II. Formalism}

\subsection*{IIa. Generalised Breit-Wigner formula for unstable
particles in a thermal bath}
 
Different species of hadrons in thermalised matter
exist with equilibrium distributions
determined by temperature $T$, the chemical potential $\mu$ and 
the statistics
obeyed by that species. Consider an unstable hadron (R) in such
a heat bath. If the decay products themselves are hadrons then they
thermalise in the bath and no distinctive decay characteristic 
can possibly be discerned out. Thus we are interested in hadrons 
(R) that decay in the heat bath (one assumes that the collision
volume can be so described) and decay into leptons and photons 
(described by the state vector $\vert f\rangle$, say) that escape 
without thermalization. With $q$ the total four momenta of the
non-hadronic final state  $\vert f\rangle$ the resonance peak  
should appear in the invariant mass ($M$) plot 
for say, the number of lepton pair events versus
$q^2=M^2$ at $M=m_R$. The mass $m_R$ of the resonance R is the mass in the 
heat bath, theoretical estimation of which will of course depend 
on the model of hadronic interactions adopted. 
The central result of Weldon's paper
is the generalization of the BW formula:
\begin{equation}
\frac{dN_f}{d^4xd^4q}=(2J+1)\frac{M^2}{4\pi^4}\frac{\Gamma_{\s{all}\ra R}
\,\,\Gamma_{R\ra f}^{\s{vac}}(M)}{(M^2-m_R^2)^2+(m_R\Gamma_{\s{tot}})^2}
\label{wel1}
\end{equation}
with $dN_f$ the number of lepton pair events, say in the space-time
and four momentum element $d^4xd^4q$, $J$ being the spin of the 
resonance, $\Gamma_{{\s{all}}\ra R}$ is the formation width,
$\Gamma_{{\s{tot}}}$ is the total width and 
$\Gamma^{\s{vac}}_{R\ra f}(M)$
is the partial decay width for off-shell R ({\it i.e.} of mass $M$) 
to go into the non-hadronic state  $\vert f\rangle$. 
While this result is deceptively 
similar to the usual BW formula, it must be realised that the thermal
distribution is implicitly contained through the `entrance' width
$\Gamma_{\s{all}\rightarrow R}$ as
\begin{equation}
\Gamma_{\s{all}\ra R}=\frac{\Gamma_{\s{tot}}}{\exp\left[\frac{}{}
\beta(E-\mu)\right]\pm 1},
\end{equation}
and also through the interpretation of $\Gamma_{\s{tot}}$ which for a bosonic
resonance R (our present concern) is given by the loss minus the gain, 
\begin{equation}
\Gamma_{\s{tot}}=\Gamma_{R\,\ra\,\s{all}} - \Gamma_{\s{all}\,\ra R},
\label{Gtot}
\end{equation}
which is actually the rate at which particles equilibrate
and relax to chemical equilibration. Here it is
important to emphasize that the width of the invariant mass plots 
is related to the thermal damping rate $\Gamma_{\s{tot}}$ and this
result generalises collision broadening treated by Van Vleck 
and Weisskopf~\cite{Van} in the context of molecular spectroscopy. 

Reverting back to the problem at hand it is helpful to have a rough
qualitative picture of decays  
for which the collision broadened BW is applicable, assuming
that the thermalised hadron fluid lasts for a time $\tau_f$ (after
which it freezes out). The amplitude for a hadron to survive at
time $t$ can be modeled, albeit noncovariantly in the frame 
of the medium (bath), by 
\ba
A(t) & = & \exp(-i\,E^\ast\,t-m_R^\ast\,\Gamma_{\s{tot}}\,t/2E^\ast)
\,\,\,\,\,\, 0<t<\tau_f\nonumber\\
& & \exp(-i\,E\,t-m_R\,\Gamma^{\s{vac}}_{\s{tot}}\,t/2E)
\,\,\,\,\,\,\,\,\,\tau_f<t<\infty
\ea
where $E$ and $m_R$ denote the energy and mass of the unstable particle
and the asterisks represent the same quantities as modified 
by the medium. Particles for which $\Gamma^{\s{vac}}_{\s{tot}}
\,\tau_f>>1$,
the relevant portion of its history is from the early period
$(0<t<\tau_f)$ as it is damped out at later times, 
and hence medium modifications determine its properties.
On the other hand for particles with 
$\Gamma^{\s{vac}}_{\s{tot}}\,\tau_f<<1$
and $1>\tau_f\Gamma_{\s{tot}}>>\tau_f\Gamma^{\s{vac}}_{\s{tot}}$ 
it is the second term that dominates and in such cases the 
thermal effects on the mass and width are negligible. 
 
Another noteworthy feature is the fact that the decay 
width $\Gamma^{\s{vac}}_{R\,\ra\,f}(M)$, to be evaluated for an 
off-shell R, occurs in the formula and not 
$\Gamma^{\s{vac}}_{R\,\ra\, f}(M=m_R)$ 
though of course the point $M=m_R$ gets weighted most 
heavily (at the peak)
due to the occurrence of the `Breit-Wigner' denominator. However, for 
broad resonances this aspect does lead to some discernible 
differences as shall be illustrated later.

Lastly, it is necessary to re-emphasize that for a particle 
which is sufficiently short lived to decay within the medium,
the width of the dilepton spectra is actually the rate at which 
it equilibrates 
($\Gamma_{\s{tot}}=\Gamma_{R\,\ra\,\s{all}} - \Gamma_{\s{all}\,\ra R}$),
involving in principle various processes in which $\s{R}$ participates.
However it may be observed that elastic scattering does not enter
into $\Gamma_{\s{tot}}$ as it cancels, contributing equally 
as it does to $\Gamma_{R\,\ra\,\s{all}}$ and 
$\Gamma_{\s{all}\,\ra R}$. This is however, in contradiction to
earlier observations made in this context~\cite{Haglin}. 
It may be borne in mind that although the elastic scattering contributes 
to the kinetic equilibrium it has no direct effect on the chemical 
equilibrium of the system while of course such elastic processes are of
importance in phenomena such as viscosity etc. Indeed elastic scattering 
changes the momentum of the colliding particles but the nature of the
particles remains unaltered and hence this process does not contribute to
the decay life time in the bath.

\subsection*{IIb. Vector Meson Dominance}

The second important element in our framework is to have a robust
phenomenological description of the electromagnetism of hadrons 
(in particular rho and omega mesons) as this shall enter into the 
decay channels of interest. Such a setting provided, for instance, by
the Vector Meson Dominance (VMD) model proposed by Nambu and
developed by Sakurai~\cite{Sakurai}, which assumes that the 
photon interacts with physical hadrons through vector mesons. 
Thus the cross
section for the process $\pi^+\,\pi^-\,\ra\,l^+\,l^-$
(where $l=e$ or $\mu$) will involve the coupling of the photon
to the pion which is expressed in terms of the pion form factor
$F_\pi(q^2)$ occurring in the matrix element of the electromagnetic
current between pion states $\langle\,\pi(p\prime)|j_\mu|\pi(p)\,\rangle
= F_\pi(q^2)(p\prime -p)_\mu$ where  the four momentum transfer is $q
=p\prime -p$. While the photon can couple directly to the pion
through its electromagnetic current $eJ_\mu = ie(\pi^- \partial_\mu
\pi^+ -\pi^+ \partial_\mu \pi^-)$, the photon can also couple to
the pion through a vector meson, which in this case, must also be
an isovector. This is taken to be the rho meson. Based on such
notions Sakurai enunciated the VMD model which has two formulations
often referred to as VMD1 and VMD2~\cite{AG}. The photon and isovector
meson part of the effective Lagrangian in the first representation is
\be
{\cal L_{VMD1}} = 
-\frac{1}{4}F^{\mn}F_{\mn}
-\frac{1}{4}\rho^{\mn}\rho_{\mn}
+\frac{1}{2}m_{\rho}^2\rho^{\mu}\rho_{\mu}
-g_{\rpp}\rho^\mu J_\mu
-eA^\mu J_\mu
-\frac{e}{2g_\rho}F^{\mn}\rho_{\mn},
\label{vmd1}
\ee
where, $\rho^{\mn}$ is the field tensor for the rho field 
constructed analogously to the
electromagnetic field tensor $F^{\mn}$, $g_{\rpp}$ may be determined
from the decay $\rho\,\ra\,\pi\,\pi$, and $g_\rho$ from fits
to the process $e^+e^-\,\ra\,\pi^+\pi^-$.
Thus here  we have a direct photon-matter coupling as well as
a photon-rho coupling which vanishes at $q^2=0$ due to the
occurrence of the derivatives in the last term above. This leads,
after provision is made for the finite width of the unstable
rho (emanating ostensibly from the imaginary part of the pion
loop in the rho self energy), to the following expression for
the rho dominated pion form factor,
\be
F_\pi^{VMD1} (q^2) = 1 - q^2\frac{g_{\rpp}}{g_\rho}
\frac{1}{q^2-m_\rho^2+im_\rho \Gamma_\rho}.
\label{fvmd1}
\ee
Note that the charge normalization constraint $F_\pi(q^2=0) = 1$
is automatically built in. Furthermore, it may be remarked that 
often a gauge-like argument is advanced to the effect that
rho couples universally (with the same strength) to all
hadrons. However, the experimental fits reveal that universality
is not exact; and indeed one finds 
\be
\frac{g_{\rpp}}{g_\rho} = 1+\epsilon,
\ee
with $\epsilon = 0.2$~\cite{Ben}. Sakurai also outlined an alternative 
formulation (VMD2), which though not as elegant as the first
(having for instance a photon mass-term in the Lagrangian), 
enjoys considerable popularity. Here 
\be
{\cal L_{VMD2}} = 
-\frac{1}{4}F^{\mn}F_{\mn}
-\frac{1}{4}\rho^{\mn}\rho_{\mn}
+\frac{1}{2}m_{\rho}^2\rho^{\mu}\rho_{\mu}
-g_{\rpp}\rho^\mu J_\mu
-\frac{e m_\rho^2}{g_\rho} \rho^\mu A_\mu 
+\frac{e^2}{2g^2_\rho} m_\rho^2 A^\mu A_\mu,
\label{vmd2}
\ee
and accordingly 
\be
\rho-\gamma ~~~~{\s {vertex}} = -\frac{i e m_\rho^2}{g_\rho}.
\label{ver1}
\ee
Furthermore 
\be
F_\pi^{VMD2}(q^2) = - \frac{m_\rho^2}{q^2 - m_\rho^2 + im_\rho\Gamma_\rho(q^2)} 
\frac{g_{\rpp}}{g_\rho},
\label{fvmd2}
\ee
where
\be
\Gamma_\rho(q^2) = \Gamma_\rho(m_\rho^2)\frac{(q^2-4m_\pi^2)^{3/2}}
{(m_\rho^2-4m_\pi^2)^{3/2}}\frac{m_\rho}{M}\Theta(q^2-4m_\pi^2),
\ee
and in order to maintain the condition $F_\pi(q^2=0) = 1$, it 
is necessary here to impose the universality condition 
viz. $g_{\rpp} = g_\rho$, whereas in VMD1 the charge normalization constraint 
is automatically maintained.
Insertion of a momentum dependent width for the unstable vector 
meson in the VMD2 form factor maintains the 
condition $F_{\pi}(q^2=0)=1$. The momentum dependence originates
on the one hand from the condition of the lowest mass state into
which the rho meson can decay namely into two pions (hence the
Heaviside theta function), and on the other hand due to the p-wave
decay (therefore, the third power of the three momentum).
We prefer VMD1 for reasons to be given
later, the relevant results for VMD2 are also discussed as this
version is used by several authors. Moving on to the isoscalar 
part of the electromagnetic interactions of hadrons this shall 
analogously be taken to be dominated by the isoscalar vector 
meson $\omega$. The relevant part of the effective Lagrangian 
density involving $\omega$ is taken to be~\cite{Sakurai,Gell}
\be
{\cal L}_\omega^{\s{relevant}} = -\frac{e m_\omega^2}{g_\omega}
\omega^\mu A_\mu + \frac{g_{\opr}}{m_\pi}
\epsilon_{\mu\nu\alpha\beta} \partial^\mu \omega^\nu \partial^
\alpha {\rho}^\beta \cdot {\pi}.
\label{wlag}
\ee
The coupling of the omega to the photon (coefficient of 
$\omega^\mu A_\mu$ above) has long been considered to be
approximately one third of that for the rho to the photon (coefficient
of $\rho^\mu A_\mu$ in Eq.(\ref{vmd2})), which yields 
reasonable agreement with the ratio of the observed partial
widths $\Gamma(\rho\,\ra\,e^+\,e^-)/\Gamma(\omega\,\ra\,
e^+\,e^-)$. Furthermore this is also supported
by a recent QCD based study~\cite{Dillon}. The coupling $g_{\opr}$ may
be determined by using this term to calculate the observed
$\omega\,\ra\,\pi^0\,\gamma$ decay through the use of the
rho-photon coupling already introduced in Eq.(\ref{ver1}).
However, the zero-ranged $\omega-3\pi$ vertex can also be 
obtained from the Lagrangian~\cite{Sakurai1},
\be
{\cal L_{\omega 3\pi}} =
f_{\omega 3\pi}\ep_{\mn\alpha\beta}\omega^\mu\epsilon^{ijk}
\partial^\nu\pi_i\partial^\alpha\pi_j\partial^\beta\pi_k,
\label{w3pi}
\ee
the latin indices referring to isospin.

\subsection*{IIc. The Walecka model and vector mesons in hot and dense
hadronic matter}

    The third ingredient needed, in order to discuss the characteristics
of the rho and omega mesons in hadronic matter at a finite 
temperature ($T$) and density ($n_B$) is to have an underlying model.
Temperatures in the range $\sim 150 - 200$ MeV and/or baryon
densities $n_B$ a few times nuclear matter density are of 
relevance. As a result the study of hadronic interactions
leading to changes in their masses and decay widths 
under such conditions assumes great significance.
Various investigations have addressed this issue over the
past several years.  
Hatsuda and collaborators~\cite{Furn}  and 
Brown~\cite{Brown} have  used the QCD sum rules  
at finite temperature and density to study the  
effective masses of the hadrons.  
Brown and Rho~\cite{Rho}  also argued that requiring chiral symmetry 
(in particular addressing the QCD
trace  anomaly)  yields  an  approximate  scaling relation between
various effective hadron masses,
which  implies,  that  all
hadronic   masses   decrease  with temperature. 
The gauged linear sigma model~\cite{Pisarski}, however, shows the opposite
trend, {\it {i.e.}} $m_\rho^\ast$ increasing with temperature.
In the present study we choose a Walecka-type
model to be delineated below to provide the medium effects on
the vector meson essentially for the sake of illustration. Many of
our remarks and subtleties shall apply in some aspects to other scenario
as well.

The relevant interaction Lagrangian in the Walecka model, 
which we have considered, and
comprising of the iso-scalar sigma, the rho, the omega
and the nucleon, is given by,
\be
{\cal L}_{\s{int}}^{\s{relevant}} = g_{\sigma}{\bar N}\phi_{\sigma}
N - g_{\rho N N}\left({\bar N}\gamma_\mu{\vec {\tau}}N - 
i\frac{\kappa_{\rho}}{2m_N}{\bar N}\sigma_{\mn}{\vec{\tau}}N
\partial^\nu\right)\cdot{\vec \rho}^\mu - g_{\omega N N}{\bar N}
\gamma_\mu N \omega^\mu.
\label{lagsw}
\ee
It may be observed that the rho has been taken to couple
both minimally and through the Pauli tensor coupling
to the nucleon. We shall adopt the value 
$m_\sigma$ = 450 MeV for the sigma mass, 
and the coupling constants shall be taken to be 
$g_{\omega N N} \sim$ 10,
$g_{\rho NN} \sim$ 2.6, $\kappa_{\rho} \sim$ 6.1, $\kappa_{\omega} =$ 0
and $g_{\sigma} \sim$ 7.4, chosen so as to reproduce the saturation
density and the binding energy per nucleon in nuclear matter~\cite{Jean}.

In the mean field approximation where the sigma meson field operator
is replaced by its (classical) ground state expectation value 
$\langle \phi_\sigma \rangle \neq 0$ the Lagrangian~(\ref{lagsw})
immediately yields a medium dependent reduction in the nucleon
mass. This is calculated from the nucleon propagator
modified by a tadpole with a nucleon loop as its head
and its tail emerging from the propagating nucleon line.
Here we also include in the Relativistic Hartree Approximation (RHA), 
the properly renormalized contribution to the baryon 
self energy from the Dirac sea as well. 
This leads~\cite{vol16,Chin} to a substantial reduction in
the nucleon mass ($m_N^\ast$) in the medium~\cite{Pradip}. It is 
in this setting that we consider the vector mesons.

To compute in-medium meson propagators one solves Dyson's equation
by essentially summing an infinite geometric series whose common
ratio is the lowest order proper polarization which comprises of
the nucleon loop (with in-medium mass $m_N^\ast$) containing both 
the particle-hole (Fermi sea) and nucleon-antinucleon (Dirac sea)
contributions. The effective mass of the vector meson ($m_V^\ast$)
in nuclear matter is obtained by finding the value of energy 
$q_0$ going to the limit $\bd q\,\ra\,0$ for which
the imaginary part of the propagator attains its maximum
or equivalently by solving the full dispersion relation 
\be
q_0^2 - {\bd q}^2- m_V^2 + {\mathrm {Re}}\Pi^D_{L(T)}(q_0,{\bd q})
 + {\mathrm{Re}}\Pi^F(q^2) = 0,
\label{eqdisp}
\ee
in the limit ${\bd q}\ra 0$ in the region where $\s{Im\,\Pi}=0$. 
$\Pi_{L(T)}$ are the longitudinal (transverse) components 
of the in-medium self energy for thermal nucleon loop.
$\Pi_F$ is the vacuum self energy of the vector meson 
with modified nucleon mass due to sigma tadpole diagram.
The expressions for $\Pi_{L(T)}$ and $\Pi_F$ are given in the appendix of 
Ref.~\cite{Pradip}.

It has been shown earlier~\cite{Pradip,gale} that the change in the rho 
mass due to rho pion interaction is negligibly small at non-zero temperature
and zero baryon density. In a different model calculation, 
it has been shown by Klingl et al~\cite{klingl} that  
to leading order in density the shift in the rho mass is very small.
Therefore the change in rho meson mass 
due to $\rho-\pi-\pi$ interaction is neglected here.
However, at non-zero density the in-medium 
modification in the spectral function of the rho meson was studied
~\cite{asakawa,friman,chanfray}
by including the medium effects in the $\rho-\pi-\pi$ vertex and 
the pion propagator in the delta-hole model. Since in this work we 
restrict our calculations within the realm of 
mean field theory (MFT) for internal lines, {\it i.e} 
the internal nucleon loop in the rho and omega self energy 
are modified due to tadpole diagram only,
the inclusion of vertex corrections and modification of the pion 
propagator due to delta-nucleon hole excitation 
will take us beyond MFT and hence are not considered here
for the sake of self consistency. Moreover, we do not include
the delta baryon in the present work, as we have simply adopted
a particular model for the sake of illustration. 

\section*{III. Dilepton Production}

The dilepton production rate due to processes occurring in
a thermalised hadronic environment is obtained by folding
the in-medium cross-section with the thermal distribution
of the participants. In this article we consider dilepton
production from pion annihilation ($\pi^+\pi^-\,\ra\,e^+e^-$)
and the rho decay ($\rho\,\ra\,e^+e^-$). 
The thermal production rate per unit four-volume for lepton pairs 
is  related to the imaginary part of the one-particle irreducible
photon self energy by~\cite{Bellac,Kapusta},
\be
\frac{dR}{d^4q}=\frac{\alpha g^{\mn} {\s{Im}}\,{\s{\bd{\Pi}}}_{\mn}(q)}
{12\pi^4q^2(e^{\beta q_0}-1)},
\label{rate}
\ee
where, $q^\mu = (q_0,{\bd{q}})$ is the four momentum of the virtual
photon and ${\s{\bd{\Pi}}}_{\mn}(q)$ is the
one particle irreducible photon self energy.
In the low invariant mass region, the pion annihilation channel is
known to be the dominant one. The invariant mass distribution of the
lepton pair in the case of pion annihilation is given by,
\be
\frac{dR}{dM}=\frac{M^3}{2\,(2\pi)^4}\,(1-4m_\pi^2/M^2)\,
\int\,M_T\,dM_T\,dy\,\sigma(q_0,{\bd q})\,\exp(-M_T\cosh\,y/T)
\label{pipigen}
\ee
Here, $q_0=M_T\cosh\,y$, ${|\bd q|}=\sqrt{q_0^2-M^2}$; 
$\sigma(q_0,{\bd q})$ is the cross-section for the pion 
annihilation calculated using VMD1 and VMD2 described by 
Eqs.~(\ref{vmd1}) and (\ref{vmd2}) with appropriate modifications
at finite temperature and density and is given by,
\be
\sigma(q_0,{\bd q}) = \frac{4\,\pi\,\alpha^2}{3\,M^2}\,\sqrt{1-4m_\pi^2/M^2}\,
\sqrt{1-4m_l^2/M^2}\,(1+2m_l^2/M^2)\,|F_\pi(q_0,{\bd q})|^2.
\ee
The $q_0$ and ${\bd q}$ dependence of the pion form factors in the
case of VMD1 and VMD2 originate from the real and imaginary part
of the rho self energy in the medium.
In the rest frame of the $\rho$ meson, Eq.~(\ref{pipigen}) reduces 
to the well known form
\be
\frac{dR}{dM} = \frac{\sigma(M)}{(2\pi)^4}\,M^4\,T\,\sum_n
\,K_1(nM/T)\,(1-4m_\pi^2/M^2),
\label{pipi0}
\ee
where $K_1$ is the modified Bessel function, and $M$ is the invariant mass
of the lepton pair. 

In the same way, the invariant mass distribution of lepton pairs from 
the decays of vector mesons propagating with a momentum ${\bd q}$ is
obtained using Eq.(\ref{wel1}), as 
\ba
\frac{dR}{dM}&=&\frac{2J+1}{2\pi^3}\,M\,\int\,M_T\,dM_T\,dy\,\exp(-M_T\cosh y)
\nonumber\\
&&\times\,\left[\frac{\omega\,\Gamma_{\s{tot}}}
{(M^2-m_V^2+{\s {Re \Pi}})^2+\omega^2\,\Gamma_{\s{tot}}^2}\right]\,
\omega\Gamma_{V\,\ra\,e^+\,e^-}^{\s{vac}}
\label{decaygen},
\ea
In the rest frame of the vector meson this reduces to
\ba
\frac{dR}{dM}&=&\frac{2J+1}{\pi^2}\,M^2T\,\sum_n\,K_1(nM/T)\nonumber\\
&&\times\,\frac{m_V^\ast\,\Gamma_{\s{tot}}
/\pi}{(M^2-m_V^{\ast 2})^2+m_V^{\ast 2}\Gamma_{\s{tot}}^2}
m_V^\ast\Gamma_{V\,\ra\,e^+\,e^-}^{\s{vac}}
\label{decay0},
\ea
where $\Gamma_{\s{tot}}$ is defined by Eq.(\ref{Gtot}),
and $\Gamma_{V\ra e^+e^-}^{\s{vac}}(M)$ is the partial width for the
leptonic decay mode for the off-shell vector particles. 

\section*{IV. Results and Discussions}

Within the ambit of the hadronic model adopted by us, the effect of
finite temperature ($T$) and density ($n_B$) on the self energies
of the vector mesons reveals that the mass of the rho meson ($m_{\rho}^\ast$)
decreases more rapidly with increasing $T$ and/or $n_B$ than that of the
omega ($m_{\omega}^\ast$) as shown in Fig~(\ref{disp}).
One observes a small difference between the longitudinal(L)
and transverse(T) modes in case of the omega meson
but in case of the rho this splitting is negligible
(attributable to the smaller vector coupling constant). 
We have observed that 
the quantity $q_0^2 - {\bd q^2}$ along the dispersion curve 
remains almost constant($\sim m_{V}^{\ast 2}$ which is defined as $q_0^2$ 
at ${\bd q = 0}$ on the mass hyperboloid). 
This means that a simple pole
approximation of the rho and omega propagator at $k^2=m_{V}^{\ast 2}$
is good enough for our calculations.  
The splitting between the transverse and longitudinal components of
the self energy of vector mesons with both vector and tensor 
interactions can be shown to be (see also ref.~\cite{Hatsuda1}),
\ba
\Pi_T-\Pi_L &=& \frac{2g_{VNN}^2}{\pi^2}\left(1-q^2(\frac{\kappa_V}{2M})^2
\right)\,\int\frac{{\bd k^2}\,d{\bd k}\,d(\cos\theta)}{\sqrt{{\bd k}^2
+M^{\ast 2}}}\left[\frac{}{}f_D+\bar f_D\right]\nonumber\\
&& \times\,\left[\frac{u\cos^2\theta-v\cos\theta+w}
{C+8k_0q_0{\bd {|k|\,|q|}}\cos\theta -
4{\bd k}^2\,{\bd q}^2\cos^2\theta}\right]
\ea
where,\\
\begin{tabular}{llll}
$u=3q_0^2{\bd k}^2-{\bd q}^2{\bd k}^2\,;$&
$v=4q_0k_0{|\bd k|}{|\bd q|}\,;$&
$w=2k_0^2{\bd q}^2+{\bd q}^2{\bd k}^2-q_0^2{\bd k}^2\,\,\&$&
$C={\bd q}^4-q_0^2k_0^2$\\
\end{tabular}
and $f_D(\bar f_D)$ is the fermi distribution for nucleons (antinucleons).
The imaginary part of the rho self energy is related to the
probability of its survival in a medium. For a rho meson propagating
in a medium with energy $\omega$ this is given by
\be
\Gamma(\omega) = \frac{g_{\rpp}^2}{48\pi}\,W^3(s)\,\frac{s}{\omega}\,
\left[1+\frac{2T}{W(s)\sqrt{\omega^2-s}}\,
\ln\left\{\frac{1-\exp[-\frac{\beta}{2}(\omega+W(s)\sqrt{\omega^2-s})]}
{1-\exp[-\frac{\beta}{2}(\omega-W(s)\sqrt{\omega^2-s})]}\right\}
\right]
\label{widthgen}
\ee
where $s = q^2 = \omega^2-{\bd q}^2$ and $W(s) = \sqrt{1-4m_{\pi}^2/s}$.
In the limit $|\bd q| \ra 0$ , the above expression reduces to 
the in-medium decay width (with the decaying particle at 
rest in the medium)
\be
\Gamma_{\rho\,\rightarrow\,\pi\,\pi} = \frac{g_{\rho\,\pi\,\pi}^2}{48\pi}\,
\omega\,W^3(\omega)
\left[\left(1+f(\frac{\omega}{2})\right)\,\left(1+f(\frac{
\omega}{2})\right)-f(\frac{\omega}{2})f(\frac{\omega}{2})
\right]
\label{width0}
\ee
It is interesting to contrast the rate given by Eq.~(\ref{widthgen})
with the one obtained purely from considerations of relativistic 
time dilation from equation~(\ref{width0}).
In this case the decay rate of an unstable particle in a general frame 
is given by,
\be
\Gamma^{\s {Lor}}_{\rho}=\frac{\sqrt s}{\omega}\,
\Gamma^{\s {rest frame}}_{\rho}
\label{widthlor}
\ee
From Fig.~(\ref{lor}) it is clear that the decay rate of a particle 
propagating in a medium with momentum $\bd k$ cannot be obtained from
its rest frame value by Lorentz transformation only. This is because
of the preferred frame of the medium.

It is well known that the naive Walecka model does not possess 
chiral symmetry, therefore, 
it is rather difficult to predict anything reliable on the pion mass. 
On the other hand, in models with chiral symmetry
such as for instance the Nambu-Jona-Lasinio model, and the linear sigma 
model with nucleons, it is  known that
the pion mass is almost unchanged as long as it is in the 
Nambu-Goldstone phase. This is simply a consequence of the
Nambu-Goldstone theorem in a medium~\cite{Hatsuda}.
So we have adopted the usual approach of keeping the
pion mass constant. 

The most significant process which contributes to the 
broadening of the omega in the thermal bath is 
$\omega\,\pi\,\lr\,\pi\,\pi$, due to which
the depletion rate of the omega increases substantially.
This issue has also been discussed by Haglin~\cite{Haglin}
where the elastic channels involving the omega were also
included. However, due to reasons mentioned earlier, we have 
only considered the above (inelastic) channel to study the
in-medium depletion of omega. Due to this process alone 
the decay rate of the omega comes out to be $\sim$ 80 MeV.
We now focus on an interesting possibility which 
becomes realisable (when 
$m_{\omega}^\ast\,>\,m_{\rho}^\ast\,+\,m_{\pi}$),
in that the decay 
$\omega\,\ra\,\rho\,\pi$, which is closed in free space 
(as $m_{\omega}\,<\,m_{\rho}\,+\,m_{\pi}$), becomes 
an open channel. 
Thus whereas the omega meson which in usual circumstances 
decays through the three particle channel $\omega\,\ra\,3\pi$, could, 
given the appropriate environment, decay by the two particle
($\rho\pi$) mode. Under these conditions the in-medium
width for $\omega\,\ra\,\rho\,\pi$ is readily found by
applying the finite temperature cutting rules to yield
\be
\Gamma_{\omega\ra\rho\pi}=\frac{g_{\opr}^2}
{32\pi m_{\omega}^3
m_{\pi}^2}\lda^{3/2}(m_{\omega}^2,m_{\rho}^2,m_{\pi}^2)
\left[\frac{}{}1+f(E_\pi) 
+f(E_\rho)\right],
\label{worp}
\ee 
where $\lambda(x,y,z)=x^2+y^2+z^2-2(xy + yz + zx)$, arises
from the phase space considerations, while
$f$ is the Bose-Einstein distribution for the
pions and the rho mesons in equilibrium. 
The coupling constant $g_{\opr}\,\sim\,2$ appearing in the
expression $\Gamma_{\omega\ra\rho\pi}$ can be 
deduced from the observed decay $\omega\,\ra\,\pi^0\,\gamma$,
using the vector dominance model of Sakurai~\cite{Sakurai}
for the $\rho\gamma$ vertex, taking the process to occur
through a virtual rho converting to the photon. The
concomitant three body process, $\omega\,\ra\,3\pi$,
is estimated from the phenomenological effective term
in the Lagrangian shown in Eq.~(\ref{w3pi}), averting
for this purpose the Gell-Mann-Sharp-Wagner~\cite{Gell} model where the 
decay proceeds through a virtual rho $\omega\,\ra\,[\rho\,\pi]\,\ra\,
\pi\,\pi\,\pi$, in order to avoid the possibility of double counting 
when the threshold for the two body decay is crossed. 
In view of the above discussion one
must include the processes $\omega\,\lr\,3\pi$, $\omega\,\lr\,\rho\,\pi$
and $ \omega\,\pi\,\ra\,\pi\,\pi$  
as these are the most important among many other possible processes
which can contribute to the broadening of omega in the medium. 
Taking all these ramifications into account the resulting width as
a function of temperature at zero baryon density and at  normal
nuclear density ($n_B^0$) is depicted in Fig.~(\ref{wrpi}).
We have noticed that the two body channel ($\omega\,\ra\,\rho\,\pi$)
opens up in the 
former case at a temperature $\sim$ 190 MeV, while in the latter
situation (normal nuclear densities)
this channel remains open even at zero temperature.
We observe a value of $\sim$ 86 MeV for  $\Gamma_\omega$ at 
a temperature of 200 MeV and normal nuclear density. 
As a result the lifetime of the omega reduces to $\sim 2.3 $ fm/c 
which should be compared to  
that of rho ($\sim 2.1$ fm/c) under the same condition. 
This is in contradiction to the commonly held
notion that the omega is too long-lived~\cite{Heinz} 
to convey any information on the fire ball in heavy ion collisions.

While pointing out a potential phenomenon interesting in 
itself, we go on to emphasize the need in general to consider
the possibility of the opening (or closing) of channels due
to the effects of finite temperature or non-zero chemical potential.
Modification of both the mass and width of the thermal omega
due to the inclusion of anomalous interactions has also
been discussed by Pisarski~\cite{Pisarski,Pisarski1}. 
\bef
\centerline{\psfig{figure=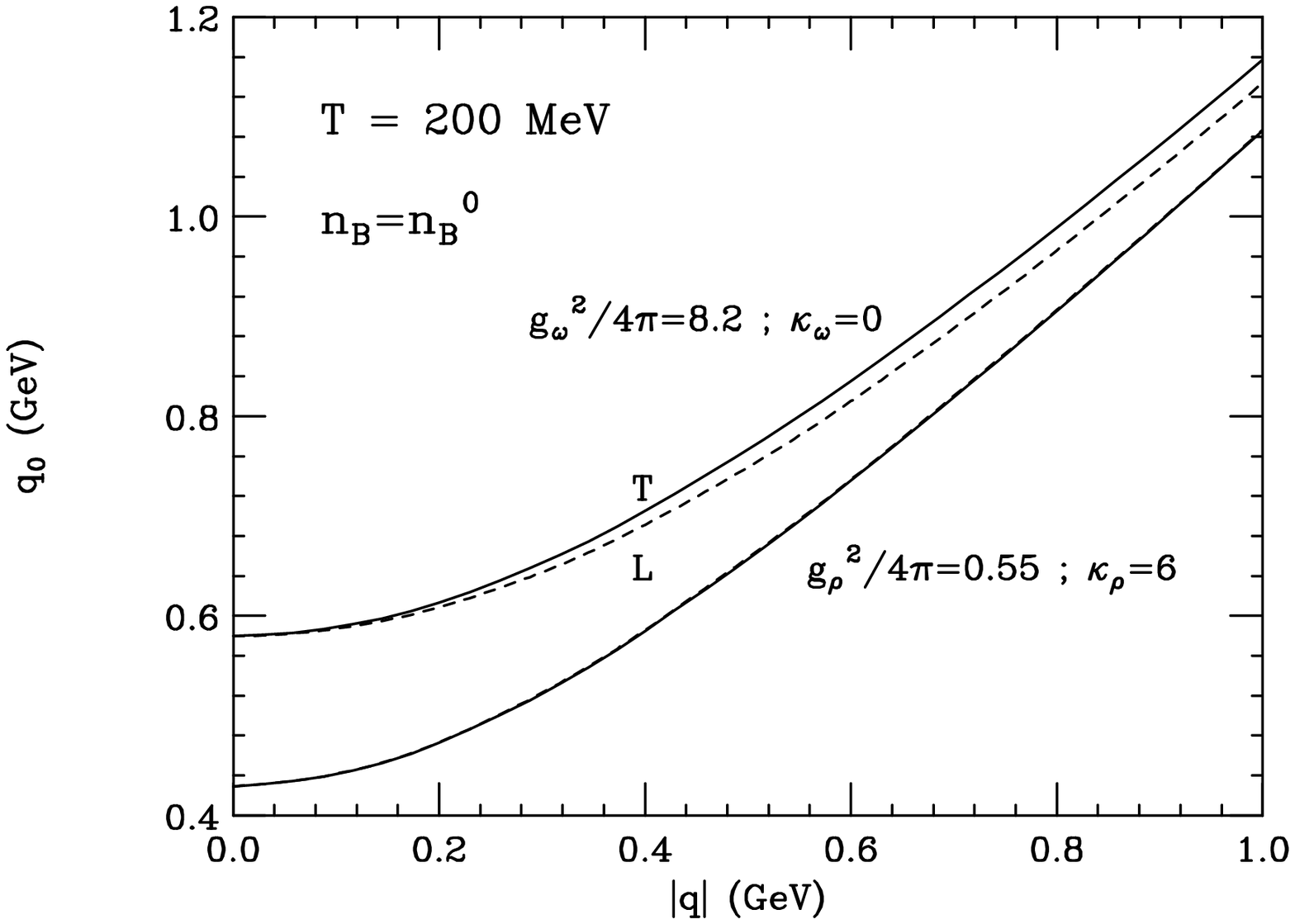,height=6cm,width=6cm}}
\caption{Transverse and longitudinal dispersion relations of rho
and omega mesons. The solid and dashed curves pertains to the 
transverse and longitudinal modes respectively.
}
\label{disp}
\eef

\bef
\centerline{\psfig{figure=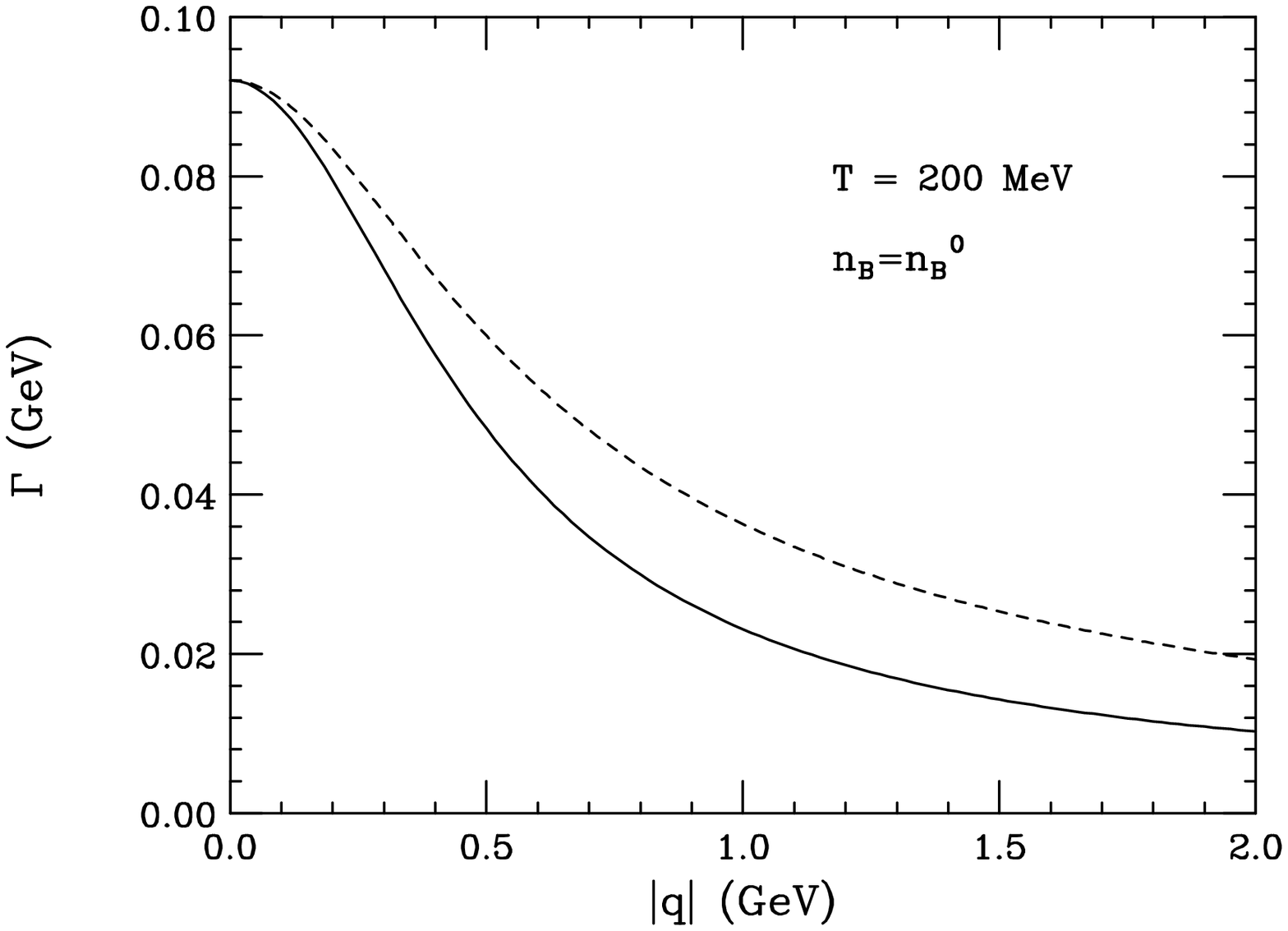,height=6cm,width=6cm}}
\caption{Decay rate of rho meson as a function of three momentum.
Solid and dashed lines correspond to the in-medium width 
[Eq.~(\protect{\ref{widthgen}})] and  the boosted width 
[Eq.~(\protect{\ref{widthlor}})] respectively.
}
\label{lor}
\eef 

\bef
\centerline{\psfig{figure=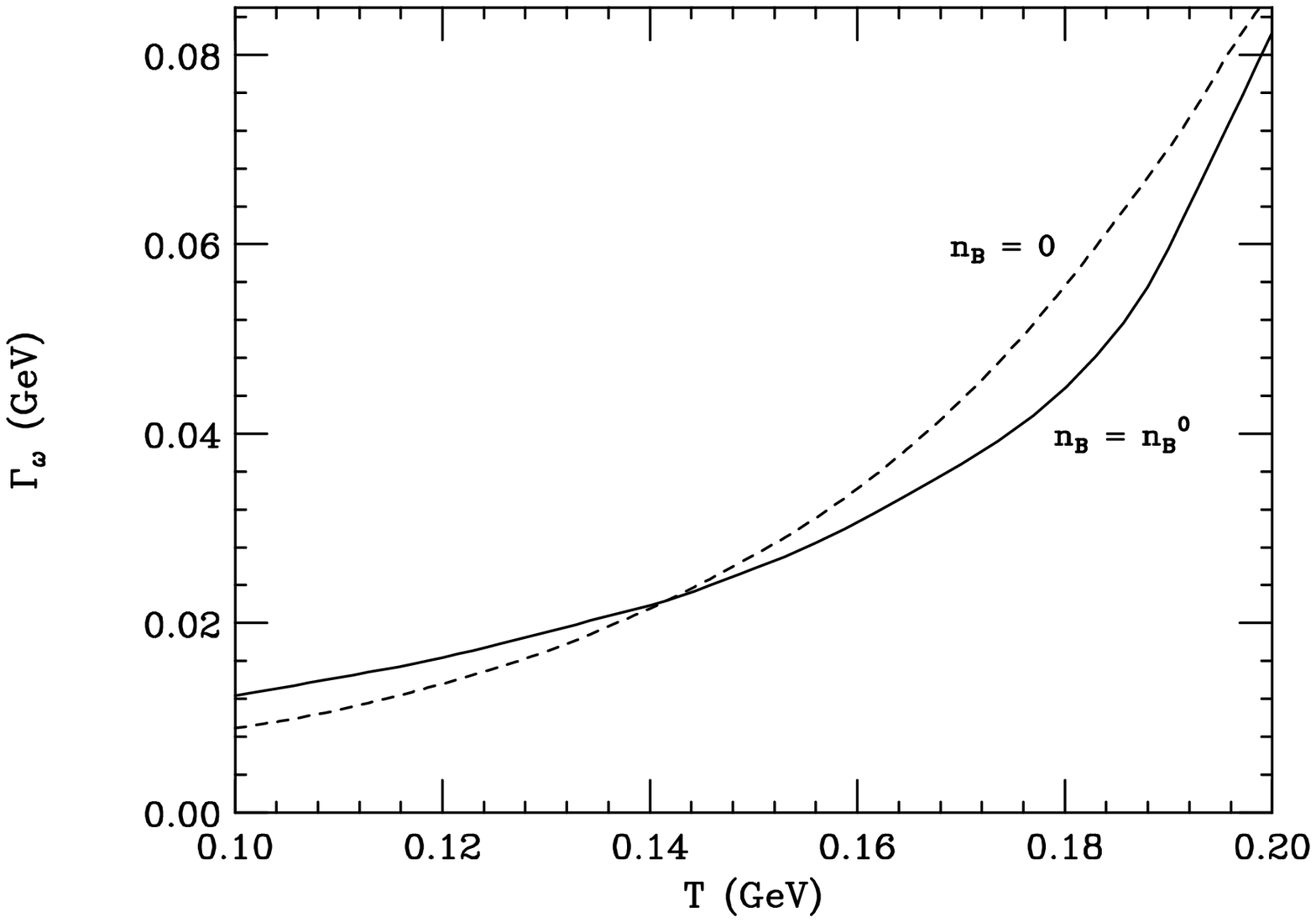,height=6cm,width=6cm}}
\caption{In-medium decay rate of omega meson comprising 
the processes $\omega\,\ra\,
\rho\,\pi$, $\omega\,\ra\,3\pi$ and
$\omega\,\pi\,\ra\,\pi\,\pi$ at $n_B = n_B^0$ (dashed line) and 
$n_B = 0$ (solid line) as a function of temperature.
}
\label{wrpi}
\eef

It has been emphasized earlier that both the forward 
and backward processes should contribute to the 
probability of propagation of an unstable particle
in a medium. This is manifested through the phase
space factors which appear in the evaluation of such a
quantity (which is neglected by some authors e.g.~\cite{Haglin}).
In this respect 
we discuss the term $[1+f(E_\pi)+f(E_\rho)]=
(1+f(E_\rho))\,(1+f(E_\pi))-f(E_\rho)f(E_\pi)$,
which occurs in the expression for the in-medium 
$\omega\,\lr\,\rho\,\pi$
width given by Eq.~(\ref{worp}). This appears
naturally in a calculation based on finite temperature field 
theory~\cite{Weldon1}. Its physical significance resides in
Bose-Enhancement (BE), which implies that the decay rate would increase
because of stimulated emission in a gas already containing the
decay products in equilibrium. 
Indeed the significant effect of this feature on
the dilepton spectra has been demonstrated in a previous 
calculation~\cite{Pradip}. This enhancement mechanism, of course,
operates also for the $\rho\,\ra\,\pi\,\pi$, $\omega\,\pi\,\ra\,\pi\,\pi$
and $\omega\,\ra\,3\pi$ decays and has been incorporated in
our calculations of decay width. 
\bef
\centerline{\psfig{figure=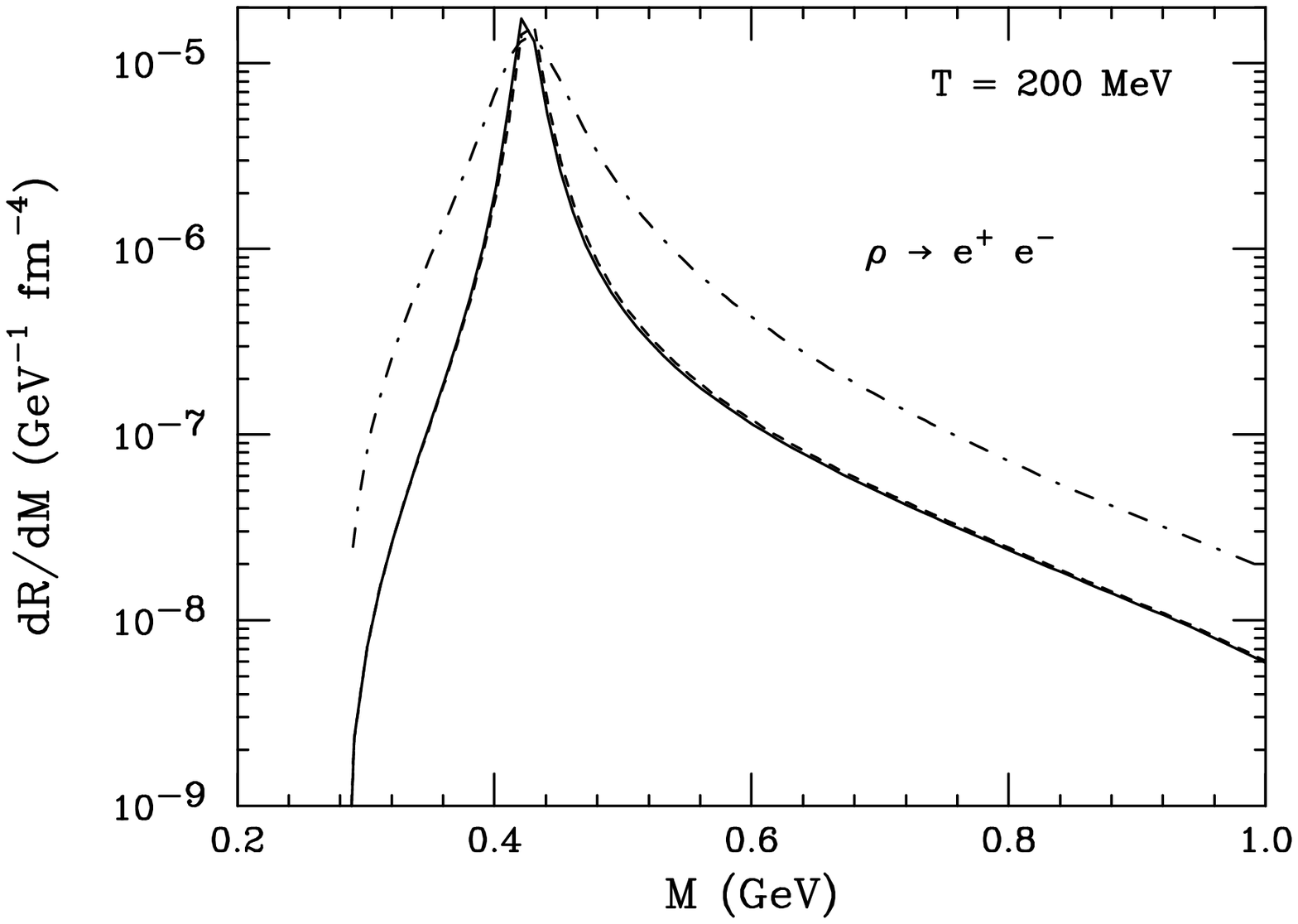,height=6cm,width=6cm}}
\caption{Invariant mass distribution of lepton pairs from
$\rho\,\ra e^+\,e^-$. The almost overlapping
solid and dashed lines correspond to transverse and longitudinal
rho with non zero three momentum. Dot-dashed line indicates the yield 
in the rest frame of the rho.}
\label{rhodec}
\eef

However, as mentioned earlier, the hadronic decay modes of rho and omega
in the fireball are not very informative and it is through their
leptonic decay modes that they become experimentally `visible'.
Therefore, it is more relevant to examine the dilepton 
emission rate from rho in the medium using Eq.~(\ref{widthgen})
which incorporates the generalised in-medium BW formula. Here we focus
our attention on $\Gamma_{\s{tot}}$. 
It is re-emphasized  that
though elastic processes such as $\rho\,\pi\,\lr\,\rho\,\pi$
are there, they do not contribute to $\Gamma_{\s{tot}}$. 
Indeed it should be borne in mind that though elastic collisions contribute
to kinetic equilibration they do not contribute to the approach
to chemical equilibrium, as indicated by $\Gamma_{\s{tot}}
\,=\,\Gamma_{R\ra\s{all}}-\Gamma_{\s{all}\ra R}$. 
For the case of the rho meson
the processes $\rho\,\lr\,\pi\,\pi$ and $\rho\,\pi\,\lr\,\omega$ are 
considered for the evaluation of  $\Gamma_{\s{tot}}$ for rho. 
From phase space considerations it is clear that
the mode $\rho\,\ra\,\pi\,\pi$ contributes dominantly to
$\Gamma_{\s{tot}}$.  The resulting  dilepton spectra from the 
decay of rho meson is shown in Fig.~(\ref{rhodec}) at $T$=200 MeV and
normal nuclear matter density. The notable feature here is the large 
shift of the rho peak towards lower invariant mass. 
This is due to the huge reduction in its mass ($m_\rho^\ast\,\sim 430 $ MeV). 
The solid and dashed curve show the resulting invariant mass distribution 
of the lepton pair from the decay of a rho propagating with momentum
$\bd q$ in the thermal bath.
The dot-dashed curve indicates the results when the decay $\rho\,\ra\,e^+e^-$
is considered in its rest frame. A broader distribution in this case
originates from the larger width of the rho in its rest frame as 
depicted in Fig.~(\ref{lor}). 
However, it has been observed that 
the effect of collisional broadening due to $\rho\,\pi\,
\lr\,\omega$ on the dilepton yield from rho decay is insignificant.

\bef
\centerline{\psfig{figure=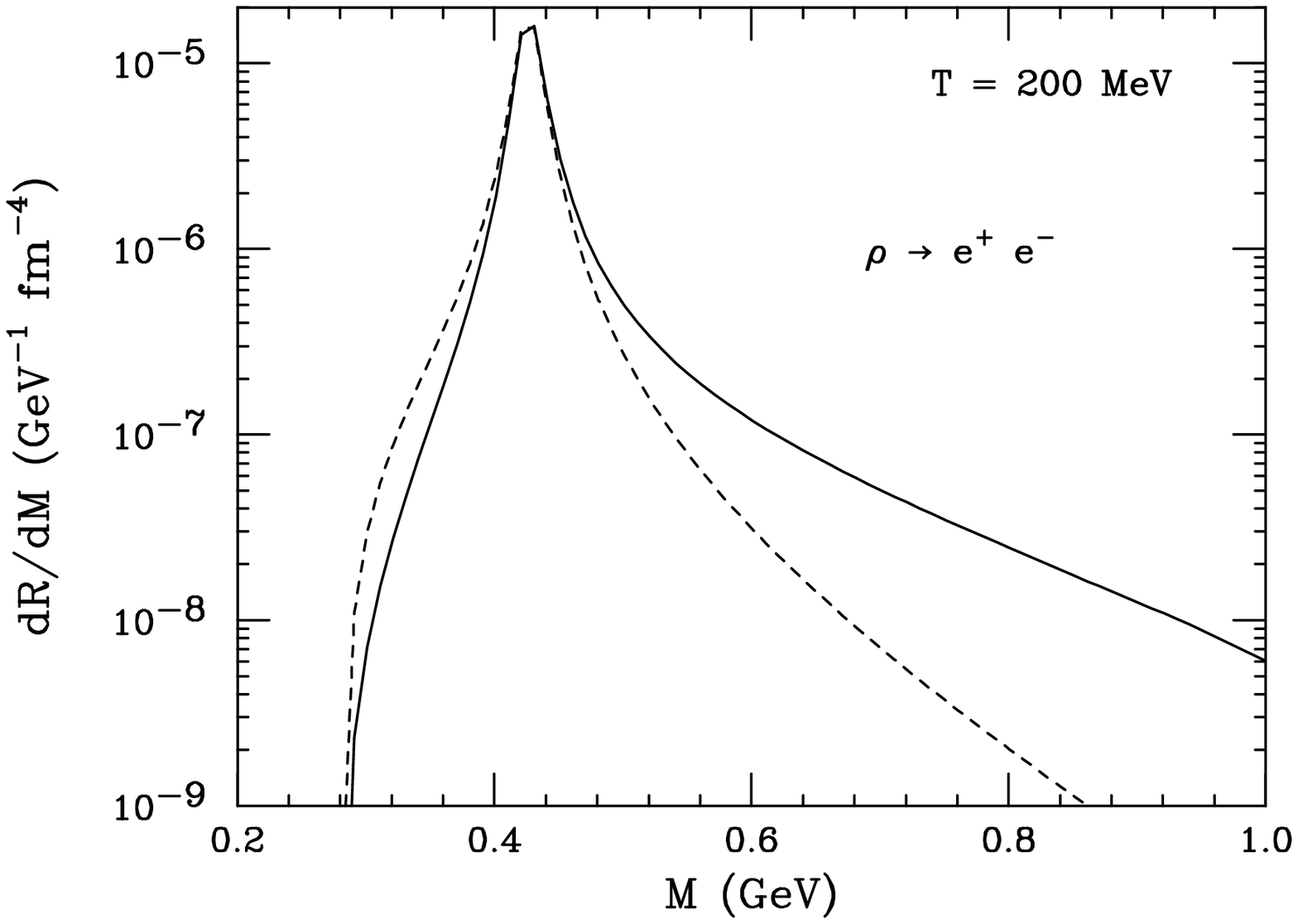,height=6cm,width=6cm}}
\caption{Invariant mass distribution of lepton pairs from
$\rho\,\ra e^+\,e^-$ at $T$ = 200 MeV and $n_B = n_B^0$.
Solid and dashed lines correspond to the case when 
$\Gamma^{\s{vac}}_{R\ra f}$ is evaluated for off-shell 
and on-shell rho (with non-zero $\bd q$) respectively.
}
\label{onoff}
\eef

We pass on to the discussion of the effect of off-shellness
of the broad rho resonance on its dilepton decay mode as exemplified 
by the occurrence of $\Gamma_{R\ra l^+l^-}^{\s{vac}}(M)$ in
Eq.~(\ref{wel1}) and again in Eq.~(\ref{decaygen})
as contrasted width $\Gamma_{R\ra\,l^+l^-}^{\s{vac}}(m_\rho^\ast)$. Of course,
in the narrow resonance limit when the BW structure reduces 
to a delta function peaked at $m_\rho^\ast$, this effect is irrelevant.
In Fig.~(\ref{onoff}) the solid curve indicates the dilepton yield
when $\Gamma_{\rho\ra e^+e^-}^{\s{vac}}$ is evaluated at $M$ (off-shell).
The on-shell result (dashed curve) shows a marked difference away from 
the rho peak (at the rho peak, of course, they must coincide). The
off-shellness in $\Gamma^{\s{vac}}_{R\,\ra\,e^+e^-}$
is calculated in the framework of VMD1. It is
relevant to remark here that the in-medium 
$\gamma-\rho$ vertex is
taken as $em_\rho^{\ast 2}/g_{\rpp}$ in VMD2 in the 
work of Li et al~\cite{Li}, which makes the coupling weaker due
to the reduction in rho mass.

\bef
\centerline{\psfig{figure=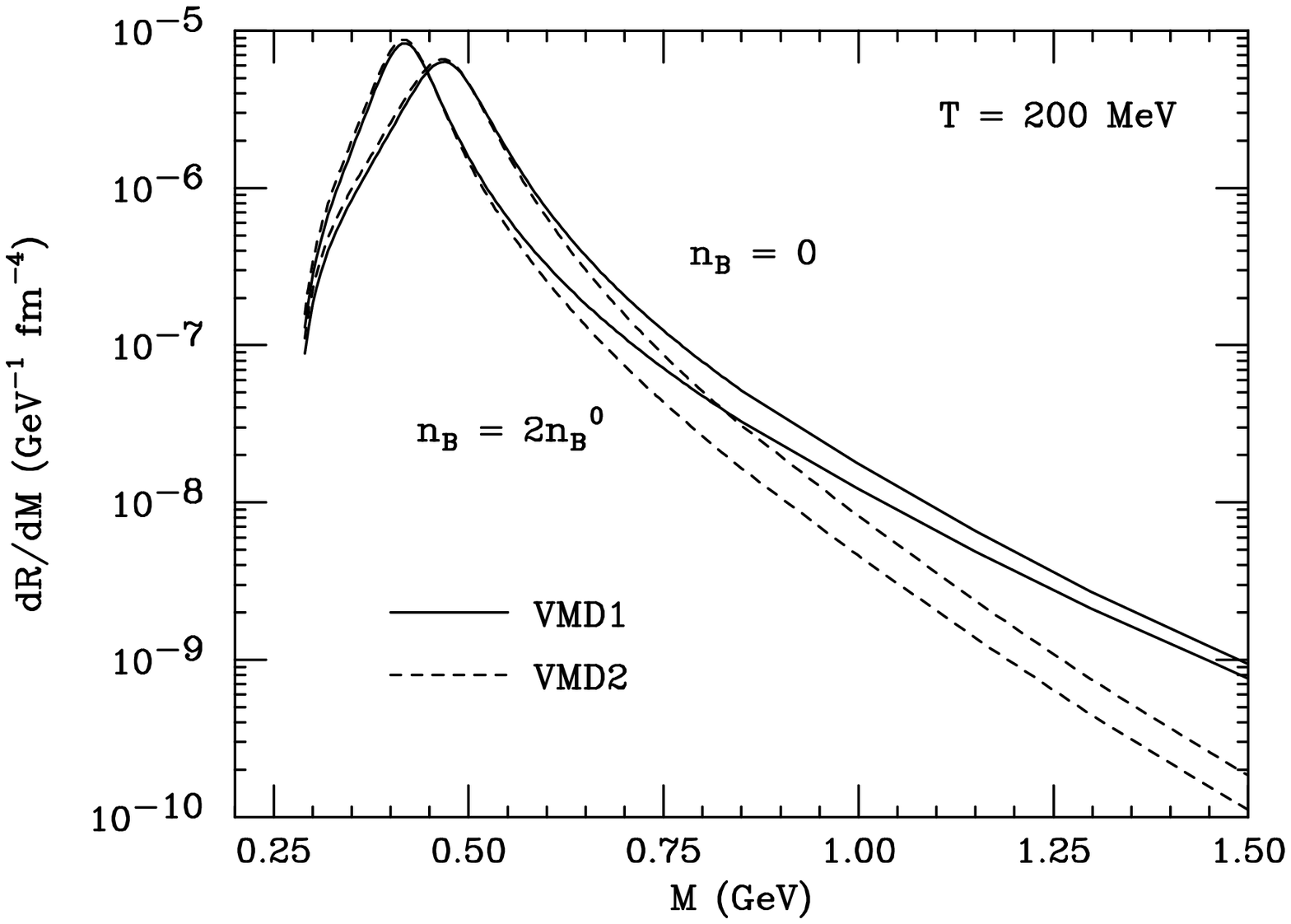,height=6cm,width=6cm}}
\caption{Invariant mass distribution of lepton pairs from
$\pi^+\,\pi^-\,\ra e^+\,e^-$ for VMD1 and VMD2.
}
\label{v1v2}
\eef

We devote this paragraph to the 
the process $\pi^+\,\pi^-\,\ra\,e^+\,e^-$, which
could proceed through a photon coupled directly
to the charge of the pion (ignoring its structure) and
for $q^2\,\neq 0$ it would begin to see its structure 
which is modeled
here by the intermediary rho meson. This separation is clearly
manifested in VMD1 as can be seen from Eq.~(\ref{fvmd1})
where the former mechanism is expressed through the occurrence
of unity and latter exhibited by the rho-pole term. 
In VMD2, however, this feature is not at all manifest.
A comparison of VMD1 and VMD2 results {\it vis-a-vis} dilepton
yield from pion annihilation at temperature 200 MeV with
$n_B=0$ and $n_B=2n_B^0$ is shown in Fig.~(\ref{v1v2})
(the momentum dependence of the rho decay width is neglected
here). The yield in the two cases is similar near the rho peak
because the form factors around $q^2=M^2\sim m_\rho^{\ast 2}$
become quite similar. However away from the peak $M>m_\rho^\ast$
the dilepton yield in the case of VMD1 dominates. For the reasons
explained here and also in the introduction we have used  
VMD1 to evaluate the dilepton yield from pion annihilation and
vector meson decays. 

Finally in order to see in what way the different issues considered
above affect the low mass dilepton spectra,
we plot in Fig.~(\ref{totdil}) the invariant mass distribution of
lepton pairs from $\rho\,\ra\,e^+\,e^-$ and
$\pi\,\pi\,\ra \,e^+e^-$. We have used VMD1 to evaluate the
contribution from pion annihilation.

\bef
\centerline{\psfig{figure=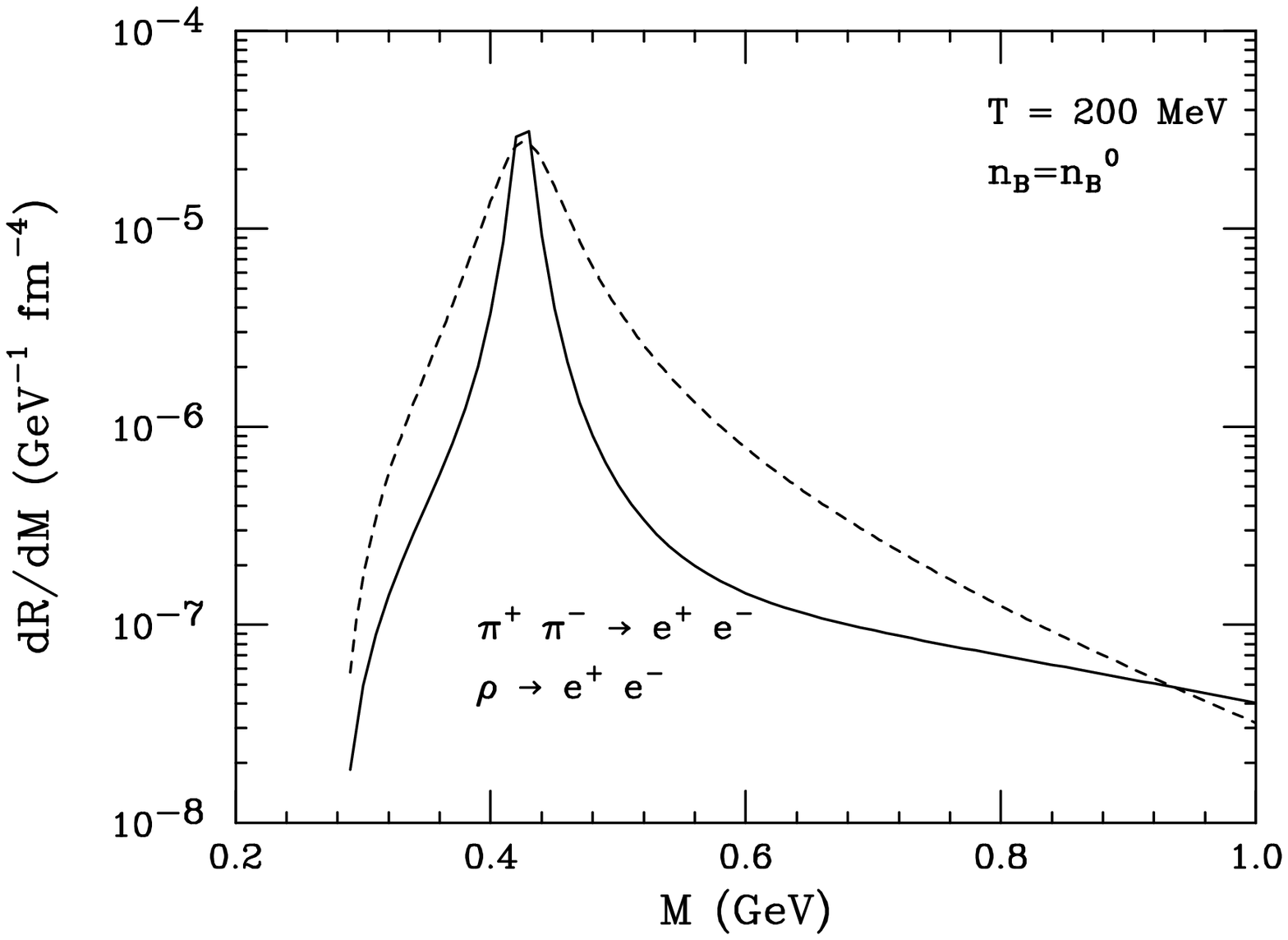,height=6cm,width=6cm}}
\caption{Invariant mass distribution of lepton pairs from
the reaction $\pi\,\pi\,\ra\,e^+\,e^-$ and the decay $\rho\,\ra\,e^+\,e^-$.
Solid (dashed) line corresponds to the case 
when energy momentum dependence of the rho self energy is
taken into account (ignored).} 
\label{totdil}
\eef

In this work we have investigated the in-medium effects
on dilepton production from pion annihilation and from the
decay of unstable particles such as the rho meson
(taken for illustration). Subtleties
arising due to the presence of a thermal bath, generalisation
of the BW formula in the medium, collisional broadening and
possibilities of double counting have been discussed.
The variation of effective masses
and decay widths of nucleons and vector mesons at non-zero
temperature and baryon density have been calculated
within the framework of Walecka model.
The BE effect in the decay width 
and the reduction in the mass of the rho meson
is found to affect the dilepton
emission rate quite substantially in the low invariant mass region.
Moreover, in the presence of nuclear matter at finite temperature a 
substantial number of omega mesons could decay inside the reaction volume and 
thus can act as a viable probe for hot hadronic matter
formed in relativistic heavy ion collisions. A comparison of the
two versions of the vector dominance model has also been presented.

Detailed measurement of photoproduction of lepton pairs 
should provide invaluable insights into the formation,
propagation and decay of vector mesons inside the nuclear medium.
Changes in the rho ( and also omega) masses would reflect directly 
in the dilepton invariant mass spectrum due to the quantum
interference between rho and omega mediated processes in the photoproduction
of lepton pairs (CEBAF Experiment). CERES collaboration ~\cite{Drees} 
has also planned to upgrade their experiment to improve the mass 
resolution such that rho and omega may be disentangled. 
Various aspects of the subtleties mentioned above 
on the observables with the inclusion of space time dynamics 
are under study, and are important for careful analysis of such
experiments.

\vskip .2in
\noindent{{\bf Acknowledgement:} We are grateful to Tetsuo
Hatsuda for useful discussions.}

\end{document}